\documentclass[twocolumn,trackchanges]{aastex62} %linenumbers

\graphicspath{{./}{figures/}}

\shorttitle{ Origin of the ridge pattern}
\shortauthors{Wang et al.}

\begin{document}

\title{Diagonal Ridge pattern of different age populations found in Gaia DR2 with LAMOST Main-Sequence-Turn-Off  and OB type Stars}
\author[0000-0001-8459-1036]{H.-F. Wang}
\affil{South$-$Western Institute for Astronomy  Research, Yunnan University, Kunming, 650500, P.\,R.\,China}
\affil{LAMOST Fellow}
\author{ Y. Huang}
\affil{South$-$Western Institute for Astronomy  Research, Yunnan University, Kunming, 650500, P.\,R.\,China}
\author{ H.-W. Zhang}
\affil{Department of Astronomy, Peking University, Beijing 100871, P.\,R.\,China}
\affil{Kavli Institute for Astronomy and Astrophysics, Peking University, Beijing 100871, P.\,R.\,China}
\author{M. L\'opez-Corredoira}
\affil{Instituto de Astrof\'\i sica de Canarias, E-38205 La Laguna, Tenerife, Spain}
\author{ W.-Y. Cui}
\affil{Department of Physics, Hebei Normal University, Shijiazhuang 050024, China}
\author{ B.-Q. Chen}
\affil{South$-$Western Institute for Astronomy  Research, Yunnan University, Kunming, 650500, P.\,R.\,China}
\author{ R. Guo}
\affil{National Astronomical Observatories, Chinese Academy of Sciences, Beijing 100101, P.\,R.\,China}
\author{ J. Chang}
\affil{Purple Mountain Observatory, the Partner Group of MPI f$\ddot{u}$r Astronomie, 2 West Beijing Road, Nanjing 210008, P.\,R.\,China}
\correspondingauthor{HFW; YH}
\email{ hfwang@bao.ac.cn {(\rm HFW)};\\
yanghuang@ynu.edu.cn {\rm (YH)}}; \\

\begin{abstract}

We revisit the diagonal ridge feature (diagonal distributions in the $R, v_{\phi}$ plane) found in $Gaia$ and present timing analysis for it between Galactocentric distances of $R=7.5$ and 12 \,kpc, using Main-Sequence-Turn-Off and OB stars selected from the LAMOST Galactic spectroscopic surveys. We recover the ridge pattern in the $R$--$v_{\phi}$ plane color coded by mean radial velocity and find this feature is presented from very young (OB stars, few hundred \,Myr) to very old populations ($\tau$ = 9$-$14 \,Gyr). Meanwhile, some ridge features are also revealed in the metallicity [Fe/H], [$\alpha$/Fe] and $v_{z}$ distributions. In the  $L_{Z}, v_{\phi}$ plane, one of the ridge patterns, with constant angular momentum per unit mass, shows variations with different age populations compared. However, the remaining two ones are relatively stable, implying there might have two kinds of ridge patterns with different dynamical origins and evolution.
\end{abstract}


\section{Introduction} 

Galactic modeling requires considering evolution and an asymmetric potential, as pointed out by \citet{antoja2018}, who have revealed in our Galaxy many intriguing signals such as snail shells, arches and ridges, etc. The so called Galactoseismology \citep{Widrow12, Widrow14} concludes the non-equilibrium and non-stationary potential of the Milky Way, observed in many density or velocity asymmetries in \citet{liuchao2017, wang2018a, wang2018b, wang2018c, wang2019, wang2020a, wang2020b, wang2020c, xu2015, Katz2018, Carrillo2019, Trick2019, Lopez2019, Lopez2020} and reference therein, which are significant for us to understand the dynamical history of the Milky Way. There is no doubt we are entering into a golden era of the Galactoseismology by embracing $Gaia$ \citep{Gaia2018} parallax and proper motions.

The inspiring snails and ridges imply that the disk is phase mixing from an out of equilibrium state \citep{antoja2018}. Hence, quadrupole patterns and phase spirals at different Galactic positions have been revealed by \citet{wangchun2019}, showing that external perturbations by the Sagittarius dwarf galaxy might be the dynamical origin for it.  Time stamps on it suggested the snails happened between 0.5 and 6 \,Gyr ago, thus leading to the consideration that young stars may have memory of the interstellar medium \citep{Tian2018}. Phase space snail shells in different cold and hot orbits distributions are also dissected in \citet{Li2020}. Unfortunately, these works using LAMOST survey \citep{Deng2012, liu2014, cui2012, zhao2012} did not investigate more details for  the intriguing ridges.

For the phase mixing patterns and structures, scenarios are mainly classified in two types: one is the external perturbations \citep{antoja2018, Binney2018, Bland-Hawthorn2019, Laporte2019, Minchev2009, Craig2019}, e.g.  Sagittarius dwarf galaxy perturbation; the other one is the internal dynamics \citep{Khoperskov2019, Barros2020, Quillen2018, Monari2019}, e.g. buckling of the stellar bar accompanied by bending waves without an external intruder. 

Both the spiral arms and Sagittarius perturbation simulations for ridges are shown in \citet{Khanna2019}. Outer Lindblad Resonance of the bar could create the prominent ridges \citep{Fragkoudi2019} and it could be used to compare with ridge map in \citet{Kawata2018}. Multiple ridges were also found in \citet{Hunt2018} with 2D transient spiral arms. Arches might be the projection of ridges in the $V_{R}$, $V_{\phi}$ plane \citep{antoja2018} and both are connected together \citep{Ramos2018}. Some recent works are also showing that the ridges could be produced by only internal mechanisms such as spirals without external contributors \citep{Barros2020, Michtchenko2019}. So far, it is still very ambiguous for us to have a clear picture for the ridges, arches, vertical waves, either the origins or relations. And whether they are from internal or external or both mechanisms is very unclear. In this work, we focus on the ridge pattern, tracing it in time stamps in a multiple-dimensional parameter space, trying to get more details of its features and better constraining its origin.There are other recent works discussing the snails, but relatively fewer works focused on ridge and without time tagging analysis as we pretend here.

The cornerstone Gaia-DR2 mission\citep{Gaia2018} has already measured precise proper motions and distances for more than 1.3 billion stars. Gaia data in combination with statistical distribution of stellar ages of millions of stars from LAMOST \citep{Deng2012, liu2014, cui2012, zhao2012} provide a good sample to study the ridge pattern, by which we can track the variation of the feature in different age populations from multiple perspectives and thus push the understanding of that without precedent in history.

This paper is organized as follows. In section 2, we introduce how we select the Main-Sequence-Turn-Off (MSTO) and OB stars sample and describe its properties concisely. The results and discussions are presented in Section 3. Finally, we conclude this work in Section 4.

\section{The Sample Selection}  

A sample of around 0.93 million Main-Sequence-Turn-Off stars with subgiant stars contribution from the LAMOST Galactic spectroscopic surveys including disk region, Galactic$-$Anticenter region, etc., is selected based on their positions in their locus in the $T_{eff}-M_{V}$ plane. With the help of LAMOST DR4 spectra and the Kernel Principal Component Analysis (KPCA) method, accuracies of radial velocities reach 5 km s$^{-1}$. The ages are determined by matching with stellar isochrones using the Yonsei-Yale (Y2) isochrones and Bayesian algorithm with the help of Teff, logg, [Fe/H] and [$\alpha$/Fe] and similar method of \citet{Jorgensen2005}. Interstellar extinction was derived using the star pairs method  and the technique is able to determine E(B−V) to an accuracy of 0.01 mag \citep{Yuan2015}, distance estimates range between 10 and 30 per cent. Overall, the sample stars have a median error of 34\% for the age estimates and the typical uncertainties of the stellar parameters, such as Teff, logg, [Fe/H] and [$\alpha$/Fe], measured from the LAMOST data are 100 K, 0.1 dex, 0.1 dex, 0.05 dex respectively \citep{xiang2017a, xiang2017b, xiang2017c}. The OB stars selection is easily selected by spectral line indices space in LAMOST and the distance here is from $Gaia$, more details could also be found in \citet{liu2019} and the data was used to unravel some velocity asymmetries in \citet{Cheng2019}.

The second data release of the $Gaia$ mission with unprecedented high-precision proper motions with typical uncertainties of 0.05, 0.2 and 1.2 mas yr$^{-1} $ for stars with $G$-band magnitudes $\leq$ 14, 17 and 20 \,mag respectively, has made possible to map the Galaxy's kinematics and Galacto-seismology with hitherto the largest spatial extent \citep{Gaia2016, Gaia2018}. 

3D velocities derived by assuming the location of Sun is $R_{\odot}$  = 8.34 \,kpc \citep{Reid14} and $Z_{\odot}$ = 27 \,pc \citep{Chen01}, \citet{Tian15} solar motion values: [$U_{\odot}$, $V_{\odot}$, $W_{\odot}$] = [9.58, 10.52, 7.01] km s$^{-1} $, other solar motions \citep[e.g., ][]{Huang2015} won't change our conclusion at all. The circular speed of the LSR is adopted as 238 km s$^{-1}$ \citep{Schonrich12} and Cartesian coordinates on the basis of coordinate transformation described in $Galpy$ \citep{Bovy2015}. We use LAMOST distance for MSTO stars by absolute magnitude and extinction measurement and Gaia parallax or distance only for OB stars. \citet{Lopez2019} have tested zero-point bias in the parallaxes of Gaia DR2, they suggested that the effect of the systematic error in the parallaxes is negligible during the work, we also don’t think the small zero point bias will affect our conclusion with similar tests to \citet{Lopez2019}. Actually, our sample are all within 4$-$5 \,kpc  away from the sun and the parallax is larger than 0.2$-$0.25 mas, the small zero point bias can not change our conclusion.

\begin{figure}
  \centering
  \includegraphics[width=0.48\textwidth]{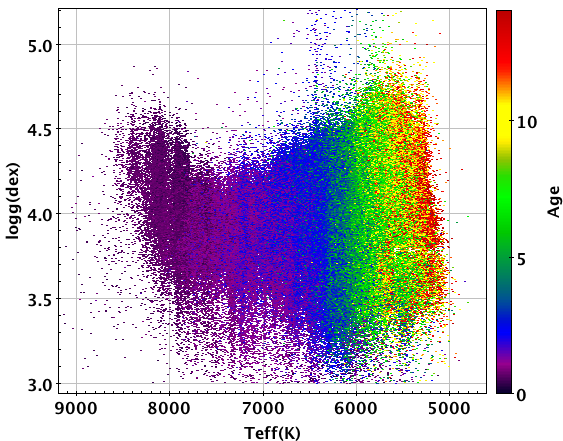}
  \caption{The figure shows the MSTO stars age distribution in the $Teff$ and $logg$ plane adopted in this work, younger stars are hotter than older stars for effective temperature.}
  \label{tefflog}
\end{figure}

\begin{figure}
  \centering
  \includegraphics[width=0.48\textwidth]{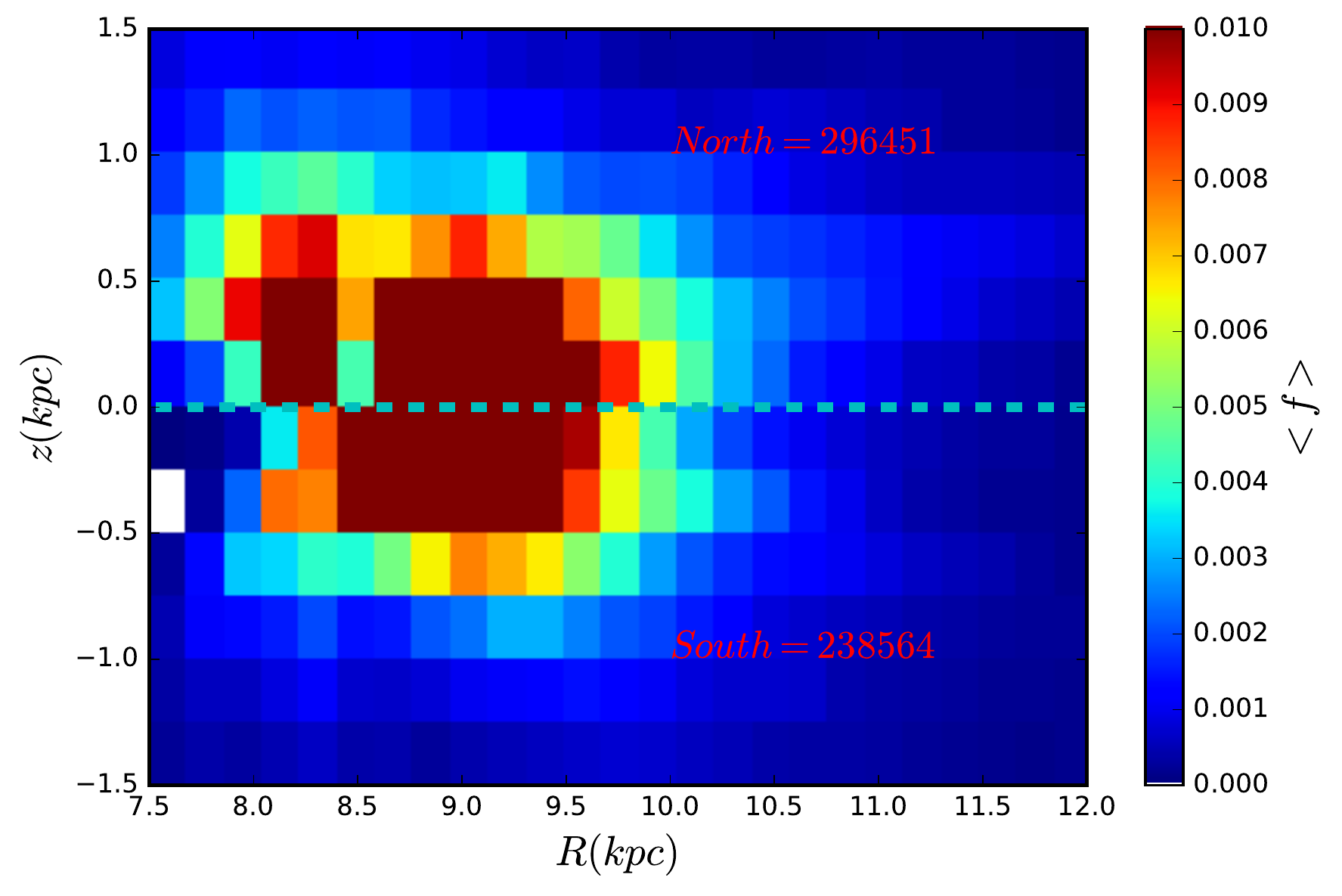}
  \caption{The figure shows the MSTO stars spatial distribution in the $R$ and $z$ plane adopted in this work, northern stars (296451) are more than southern stars (238564) by comparing the number of both sides.}
  \label{ridge_NS_mapcount1}
\end{figure}

We show the MSTO sample in Fig. \ref{tefflog}. It shows the $Teff$ vs. $logg$ distributions colored by age, we can see most of stars have surface gravity larger than 3, and younger stars have higher effective temperature than the old ones. The Fig. \ref{ridge_NS_mapcount1} shows the star counts distribution in the $R, Z$ plane, it shows that the northern stars is more than southern stars with the calculations. In order to build the reliable sample containing stellar astrophysical parameters and precise kinematical information, we use criteria from LAMOST spectroscopic survey and Gaia catalogs as follows:  

1) $|Z|$ $<$ 1.5 \,kpc and 7.5 $<$ $R$ $<$ 12 \,kpc; 

2) SNR $>$ 20; 

3) age less than 14 \,Gyr and larger than 0; 

4) $v_{\phi}$ = [50, 350] km s$^{-1}$; 

5) parallax $>$ 0 and the relative error $<$ 0.20. 

\section{Results and Discussions} 

\subsection{Ridge patterns investigation for MSTO stars}

\begin{figure*}[!t]
\centering
\includegraphics[width=0.307\textwidth, trim=0.0cm 0.0cm 0.0cm 0.0cm, clip]{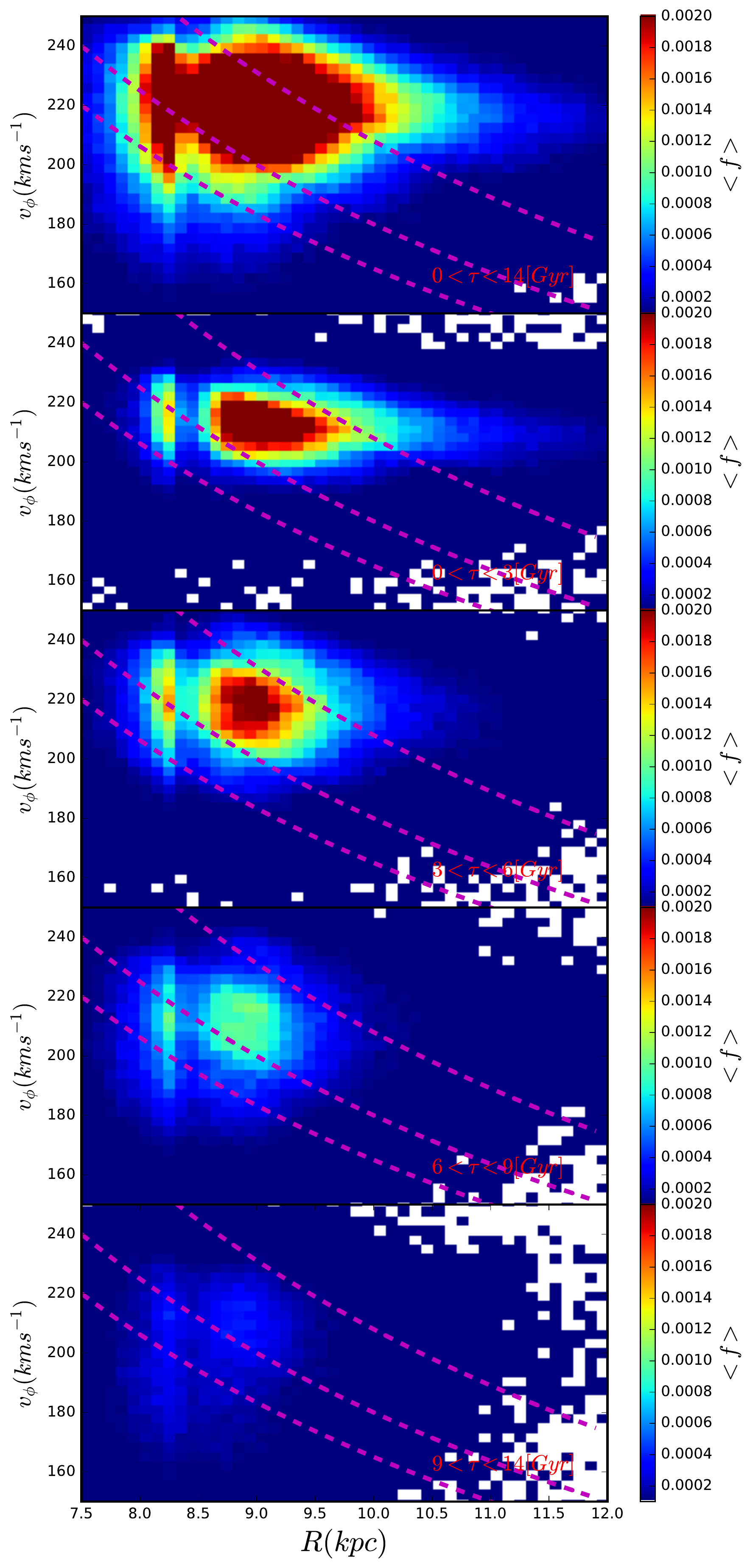}
\includegraphics[width=0.3\textwidth, trim=0.0cm 0.0cm 0.0cm 0.0cm, clip]{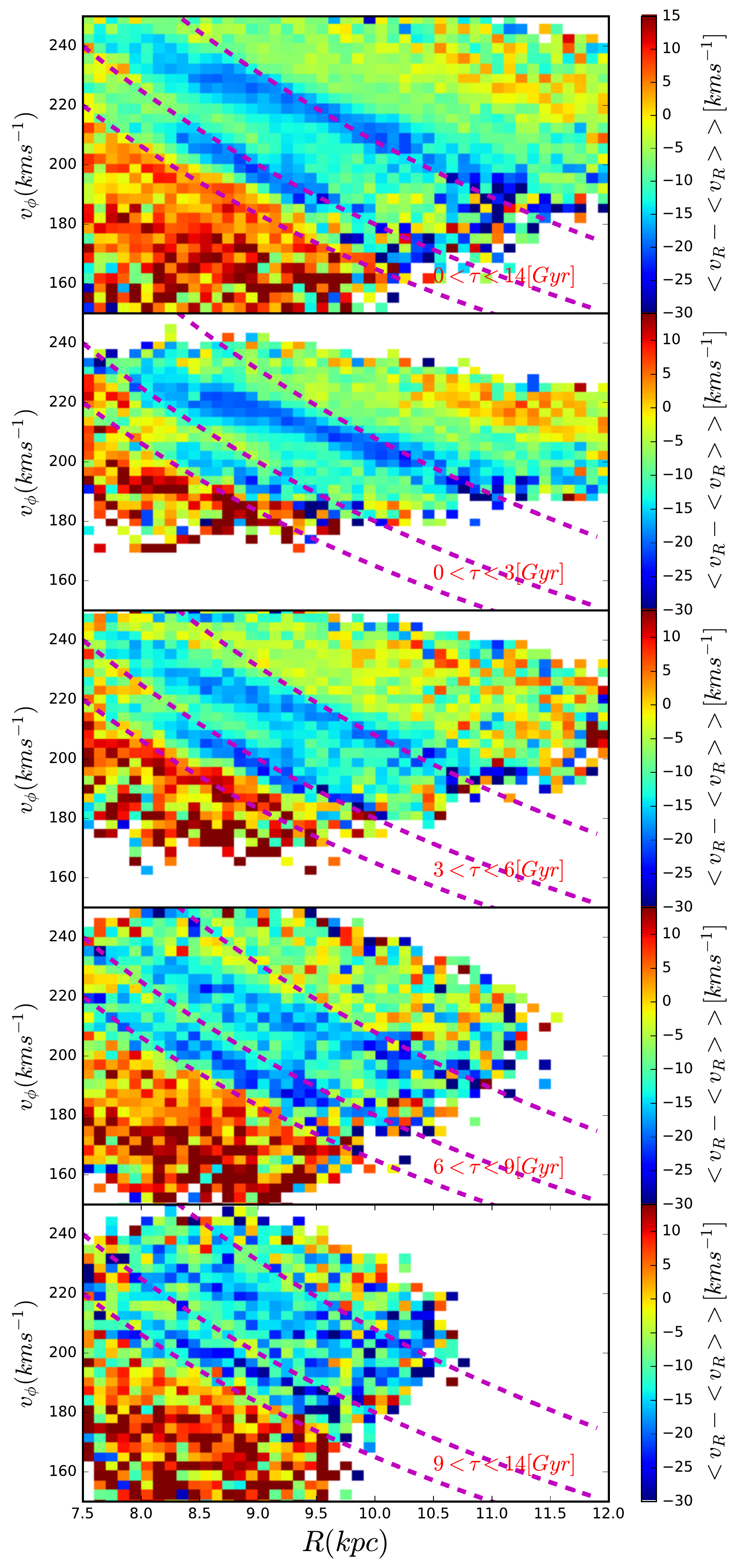}
\includegraphics[width=0.3\textwidth, trim=0.0cm 0.0cm 0.0cm 0.0cm, clip]{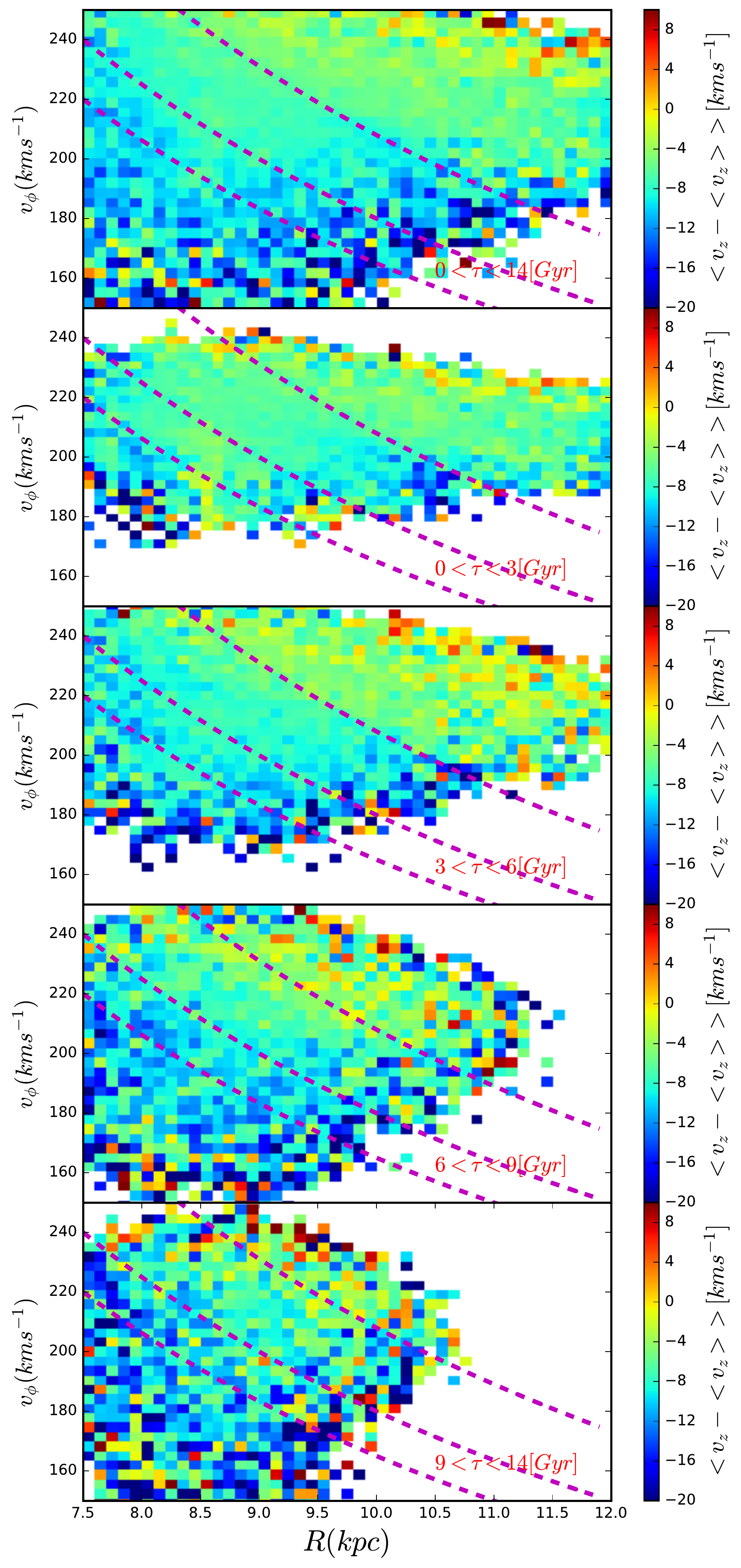}
\caption{Stars distribution in the ($R, v_{\phi}$) plane with LAMOST MSTO stars and Gaia DR2 proper motion in different age populations. Heat maps of various quantities are shown; Left panel is the density $f$ distribution, middle one is the radial motion $v_{R}$, the right panel is the vertical velocity $v_{z}$. The magenta dotted curves represent constant angular momentum of $L_{Z}$ = (1650, 1800, 2080) \,kpc km $s^{-1}$ with contribution of the $V_{LSR}$. The radial distance range is from 7.5 to 12 \,kpc.} 
\label{msden_vr_feh}
\end{figure*}

For this part, we investigate the ridge pattern in the different parameter space plots. We show in Fig. \ref{msden_vr_feh} density ($f$), radial velocity ($v_{R}$) and vertical velocity ($v_{z}$) distributions in the plane of the rotational velocity in the y axis and radial distance in the x axis; the magenta dotted curves represent constant angular momentum of $L_{Z}$ = (1650, 1800, 2080) kpc km $s^{-1}$ including the contribution of the $V_{LSR}$ \citep{Schonrich12}. The radial distance range displayed here is from 7.5 to 12 \,kpc. We can see  there are no clear ridge features for the density pattern  due to the selection effects, sample precision, etc. However, the ridge pattern is very prominent and strong for the radial velocity distribution shown in the middle sub-figures, shown as negative blue strips accompanied with positive red stripes, patterns and trends of which are similar with some previous works. e.g., \citet{Khanna2019}. In addition, it denotes clearly that the sensitive time of the inspiring ridge feature to the possible perturbation is 0$-$14 \,Gyr due to that we could detect the ridge signals in the range of 0 to 14 \,Gyr.

It is remarkable that the angular momentum per unit mass of the ridge pattern varies with age when compared with the constant magenta  lines in radial velocity sub-figures showing in the middle one. As we can see in the top panel of middle figure of Fig. \ref{msden_vr_feh}, there are three magenta lines we adopted corresponding to the three ridges colored by blue and red strips. Here we define these three strips as ridge A, ridge B, ridge C from the top to the bottom. For the ridge A in the top, we could see when the age is less than 6 \,Gyr. The pattern could be matched with the constant angular momentum line for the general trend by focusing on the range of 9$-$10.5 \,kpc, but  when we move forward with age larger than 6 \,Gyr, the ridge pattern, especially for the range of 9$-$10.5 \,kpc, is deviating the magenta lines with around 10 km s$^{-1}$ in all population bins. The corresponding errors are small and almost all of them are less than 2$-$4 km s$^{-1}$. For the ridge B, the overall trend of the ridge pattern could be matched with the second magenta lines in all populations without significant deviations, implying that there is no significant variation compared with the constant angular momentum line. For the ridge C, the minimum one, it is also matched with the third magenta lines well and has no clear variation like ridge A by focusing on the range of 9$-$10.5 \,kpc to guide our eyes. We suggest the relatively variable ridge A ($\tau > $ 6 vs. $\tau < $ 6 \,Gyr) and relatively invariable ridge B, C are showing two kinds of ridges possibly originated from different physical scenarios, which is helpful for us to unveil the origins of the ridge.

\begin{figure}
  \centering
  \includegraphics[width=0.48\textwidth]{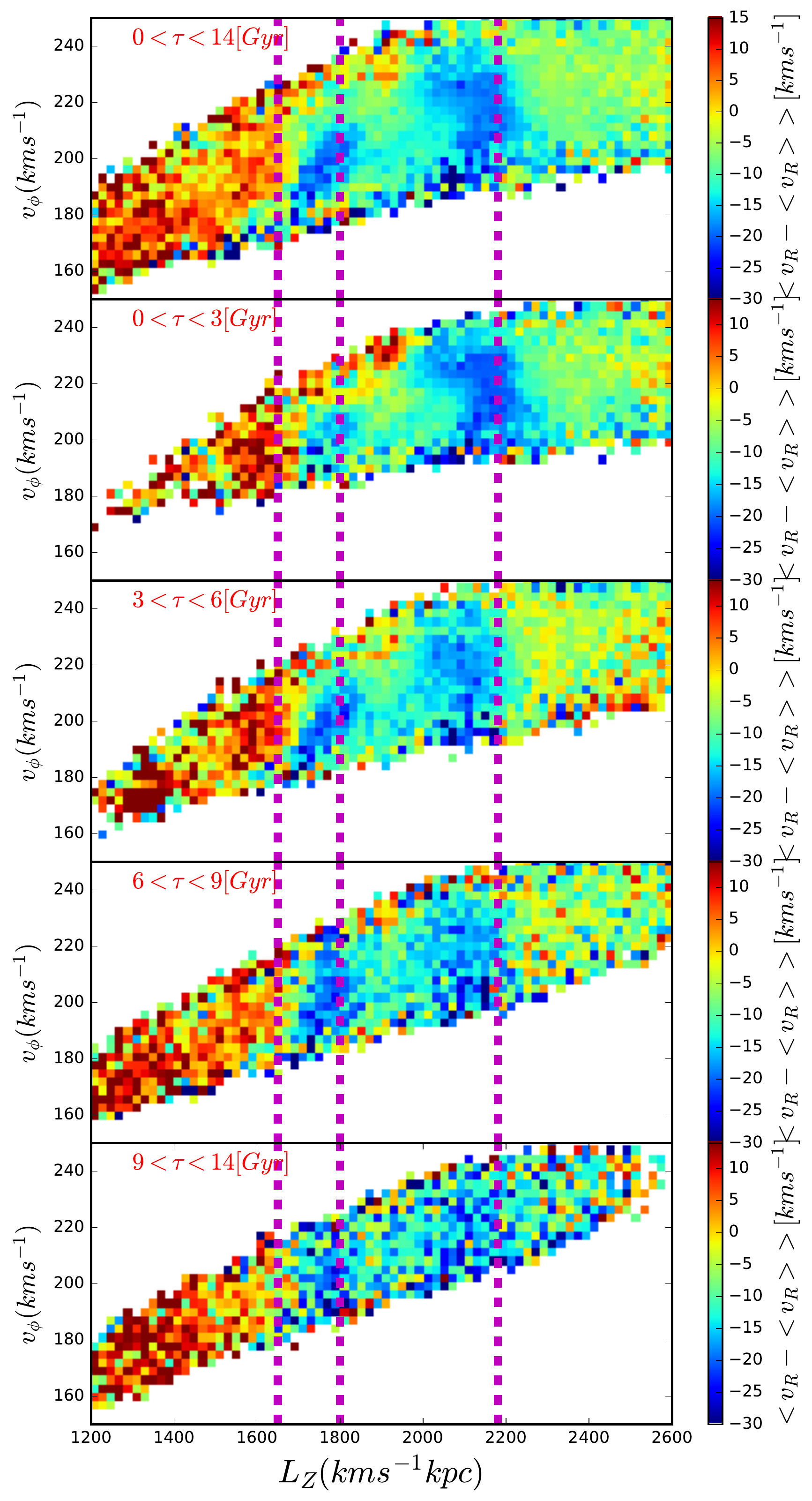}
  \caption{The figure shows the MSTO stars radial velocity distribution in the $v_{\phi}$ and $L_{Z}$ plane adopted in this work, there are three ridges around $L_{Z}$ = (1650 1800 2180) \,kpc km $s^{-1}$ plotted by the vertical magenta lines, similar to the Fig. \ref{msden_vr_feh} for the left two lines,  please notice the right line used here has small difference from the Fig. \ref{msden_vr_feh} in order to guide our eyes clearly.}
  \label{ridge_Lz_VR}
\end{figure}

So there are two relatively stable ridges and one variable ridges from the current figures and in order to try to see these ridges variation clearly,  according to \citet{Friske2019}, in Fig. \ref{ridge_Lz_VR} we also use the $L_{Z}$ vs. $v_{\phi}$ panel colored by $v_{R}$  to investigate the ridge pattern, we could see there are also three ridge patterns from left to the right and colored by red, blue and blue. These ridge patterns are corresponding to the three ridges of the Fig. \ref{msden_vr_feh}, we can see the left two ridges are relatively stable but the shift of the right ridge located around 2180 \,kpc km $s^{-1}$ of $L_{Z}$ are detected, especially when the age is larger than 3 \,Gyr. Please notice that here we use 2180 but not 2080 vertical line for $L_{Z}$ due to that the angular momentum must have larger errors than the radial distance with the contribution of velocity and distance errors for $L_{Z}$, so the pattern here has some differences from the Fig. \ref{msden_vr_feh}, we use relatively larger value for the right vertical magenta line in order to see the variable pattern in the right relatively clearly and try to match the pattern of the Fig. \ref{msden_vr_feh}, which could not affect our conclusions. \citet{Friske2019} suggested the $L_{Z}$  shift of resonances with the age is expected, because of higher energy E (or action J) of the older stars. We also propose that the shift of the ridge again supports us that there might have two kinds of ridges.

When we keep going with the $v_{z}$ pattern in the same plane, just as shown in the right panel of Fig. \ref{msden_vr_feh}, what we could see is that weak ridge features are observed in the vertical motions especially around the bottom magenta line, although it is not as strong as radial velocity, which is also not as clear as the results of \citet{Khanna2019} showing clear pattern in the vertical velocity distribution. The ridge stars in \citet{Khanna2019} are mainly consisting of mid-plane stars less than 0.2 \,kpc, which is different from our results here using stars less than 1.5 \,kpc (in order to get more stars and see the ridge pattern in $v_{R}$ clearly). When we use similar but more stringent selection conditions, the sample is too small to see very clear pattern like Fig. \ref{msden_vr_feh} due to the increasing poisson noise and observational errors, etc.

\begin{figure*}[!t]
\centering
\includegraphics[width=0.3\textwidth, trim=0.0cm 0.0cm 0.0cm 0.0cm, clip]{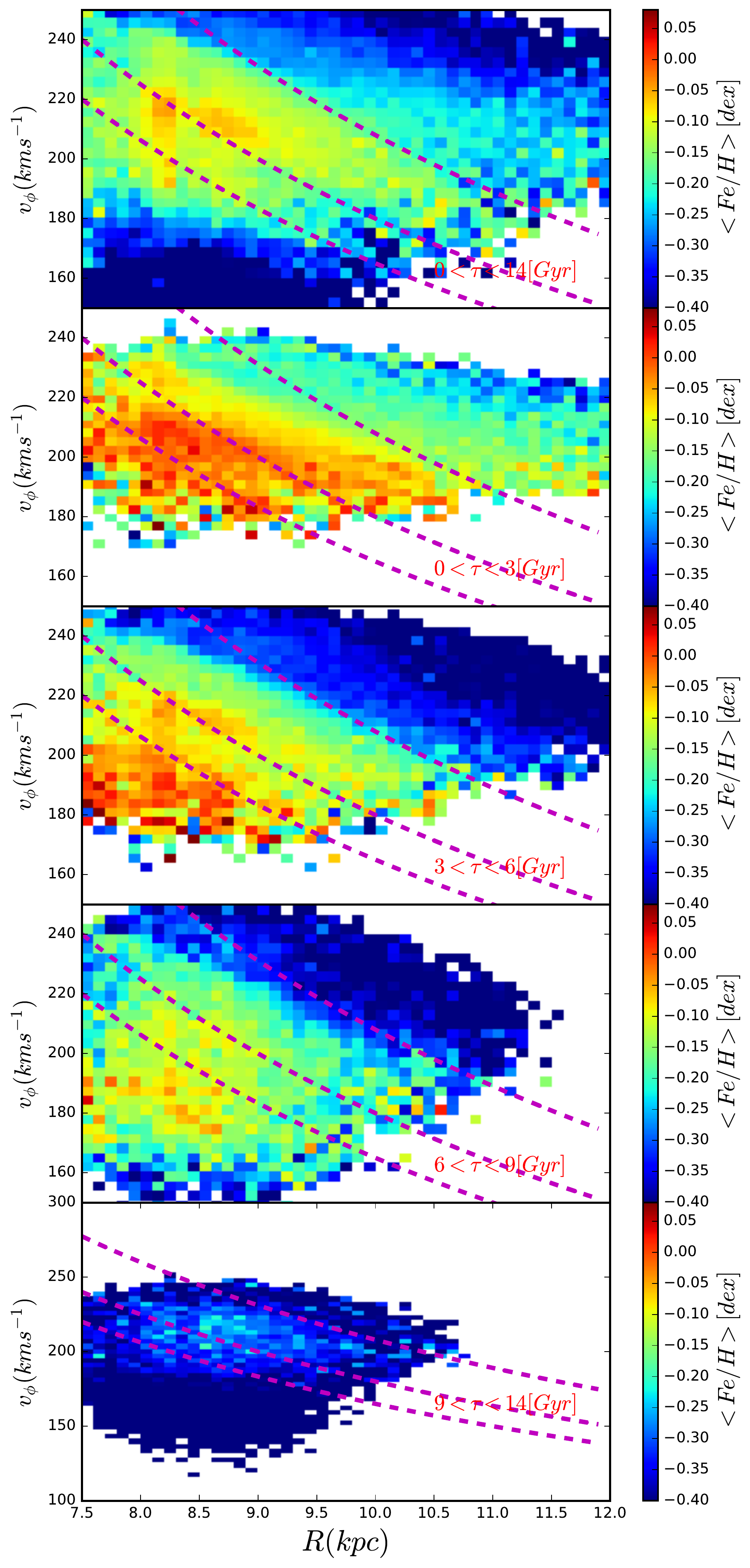}
\includegraphics[width=0.3\textwidth, trim=0.0cm 0.0cm 0.0cm 0.0cm, clip]{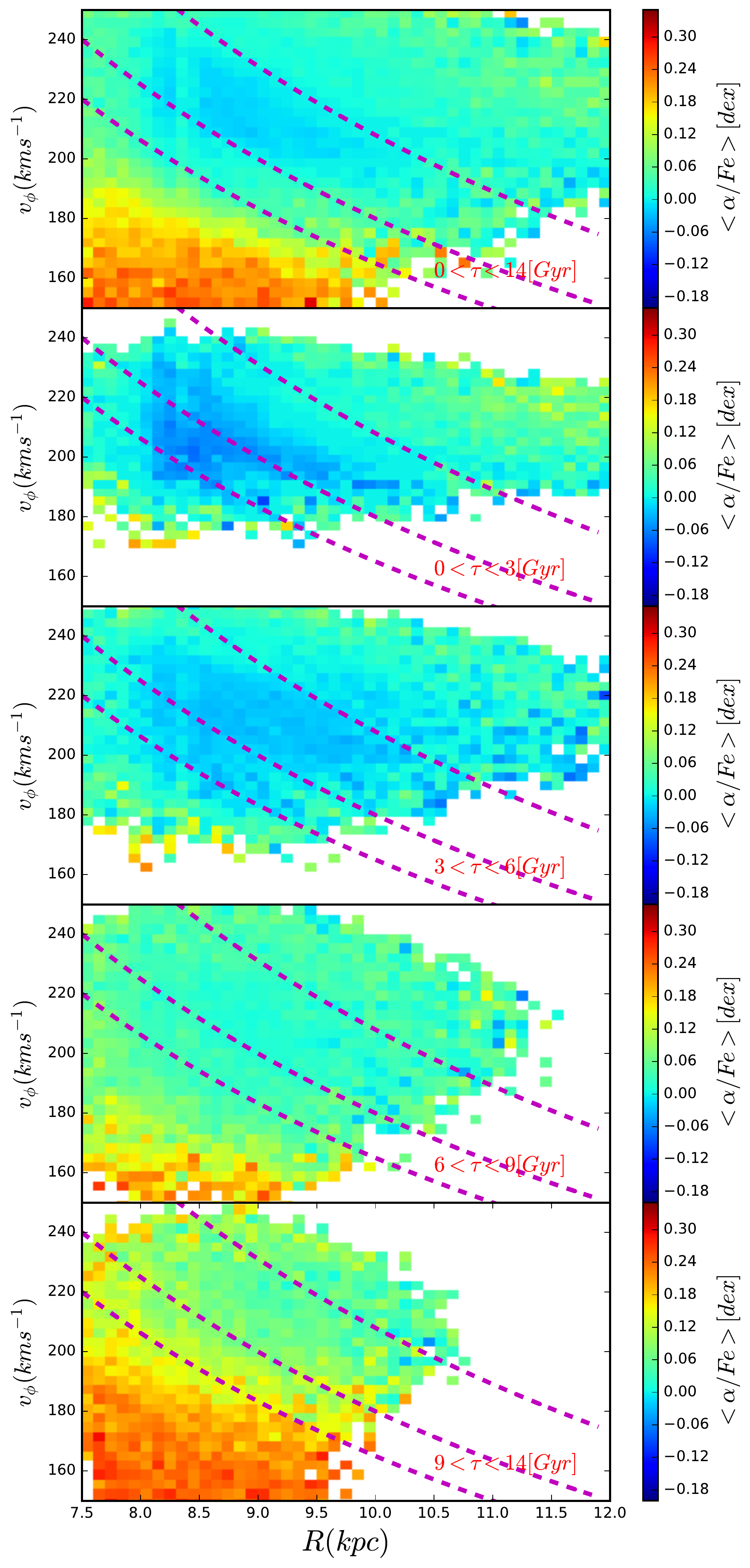}
\includegraphics[width=0.295\textwidth, trim=0.0cm 0.0cm 0.0cm 0.0cm, clip]{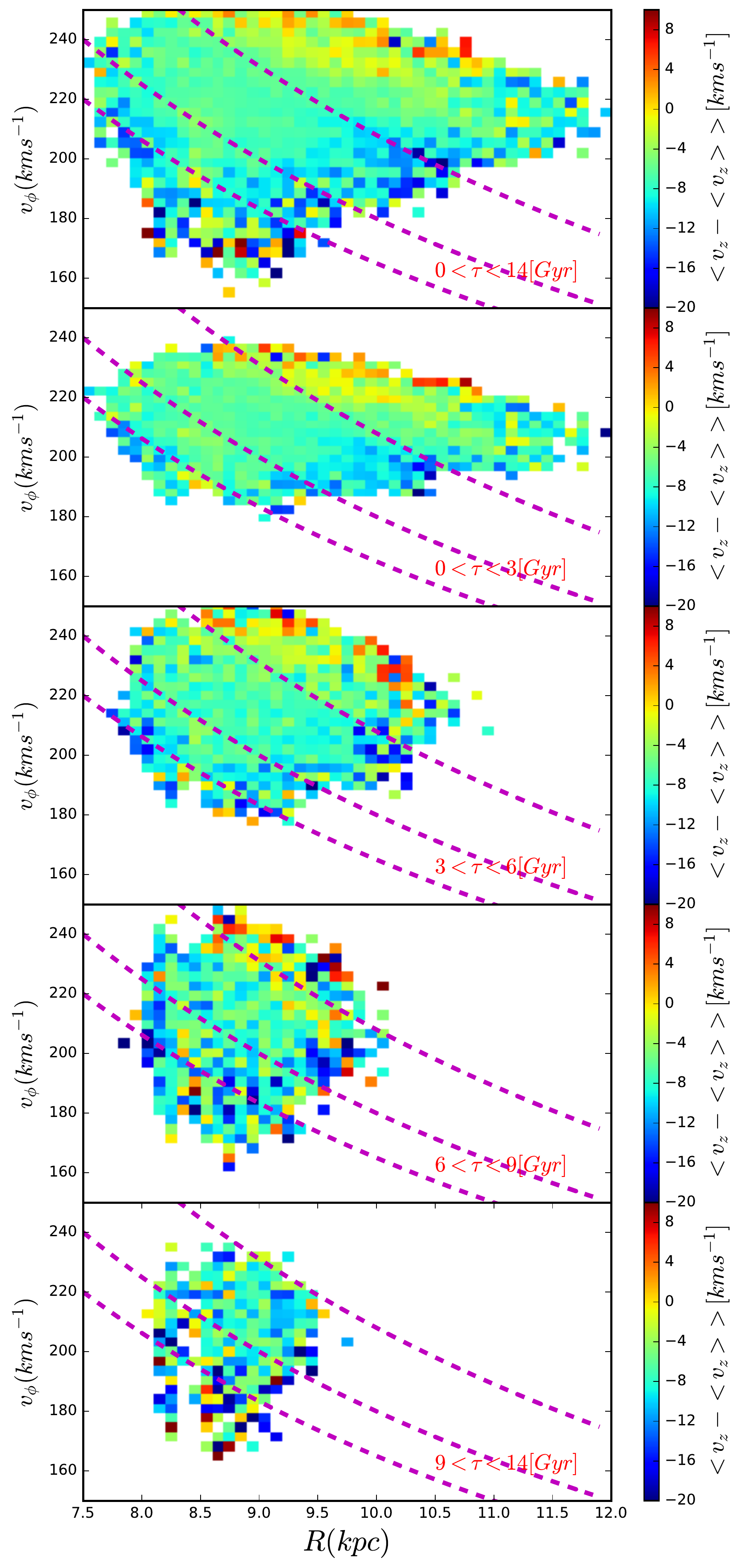}
\caption{The chemical [$Fe/H$], $\alpha$-abundance belonging to z = [$-$1.5, 1.5] \,kpc and vertical velocity in the range of  z = [$-$0.2, 0.2] \,kpc, distributions in the rotational velocity (y) and radial distance(x) plane are shown in the left, middle and right panel respectively. Ridges are detected in the left two figures and weak ridge signal is shown in the right figure.} 
\label{msden_alpha_vz}
\end{figure*}

All stars with all heights contribute to the ridge but in this case, the vertical information might be completely washed out  \citep{Khanna2019}. We could test whether there is a possibility that these factors mentioned here and last paragraph also might affect the distribution of [$Fe/H$], [$\alpha/Fe$], $v_{z}$. As shown in Fig. \ref{msden_alpha_vz}, the [$Fe/H$], [$\alpha/Fe$] (z=[$-$1.5 1.5] \,kpc), and $v_{z}$ (z=[$-$0.2 0.2] \,kpc) distributions are displayed that there are still weak ridge features in the metallicity and abundance, especially for the top panels in the left and middle figure. We have plotted there red and blue strips in the range of 8$-$10 \,kpc and around 220 km $s^{-1}$. Other features are not so clear, but we could still detect some signals, e.g., the third row of the left panel and the second row of the middle figure. By  using a narrow range of stars, we could detect weaker signal around the bottom magenta line for the vertical distribution in the right panel of  Fig. \ref{msden_alpha_vz}.

Summing up, in the $v_{\phi}, R$ plane for our sample, we see the well-known ridge pattern in radial velocity accompanied by the signals in the [$Fe/H$], [$\alpha/Fe$] and $v_{z}$ distributions, with the time tagging analysis not shown in previous works yet. They are displaying observational evidence that the different ridges might have different angular momentum that is variable or not with time, which shows for the first time there are possibly two types of ridges with different properties and origins. 

\subsection{Ridge patterns investigation by OB stars}

\begin{figure*}[!t]
\centering
\includegraphics[width=0.46\textwidth, trim=0.0cm 0.0cm 0.0cm 0.0cm, clip]{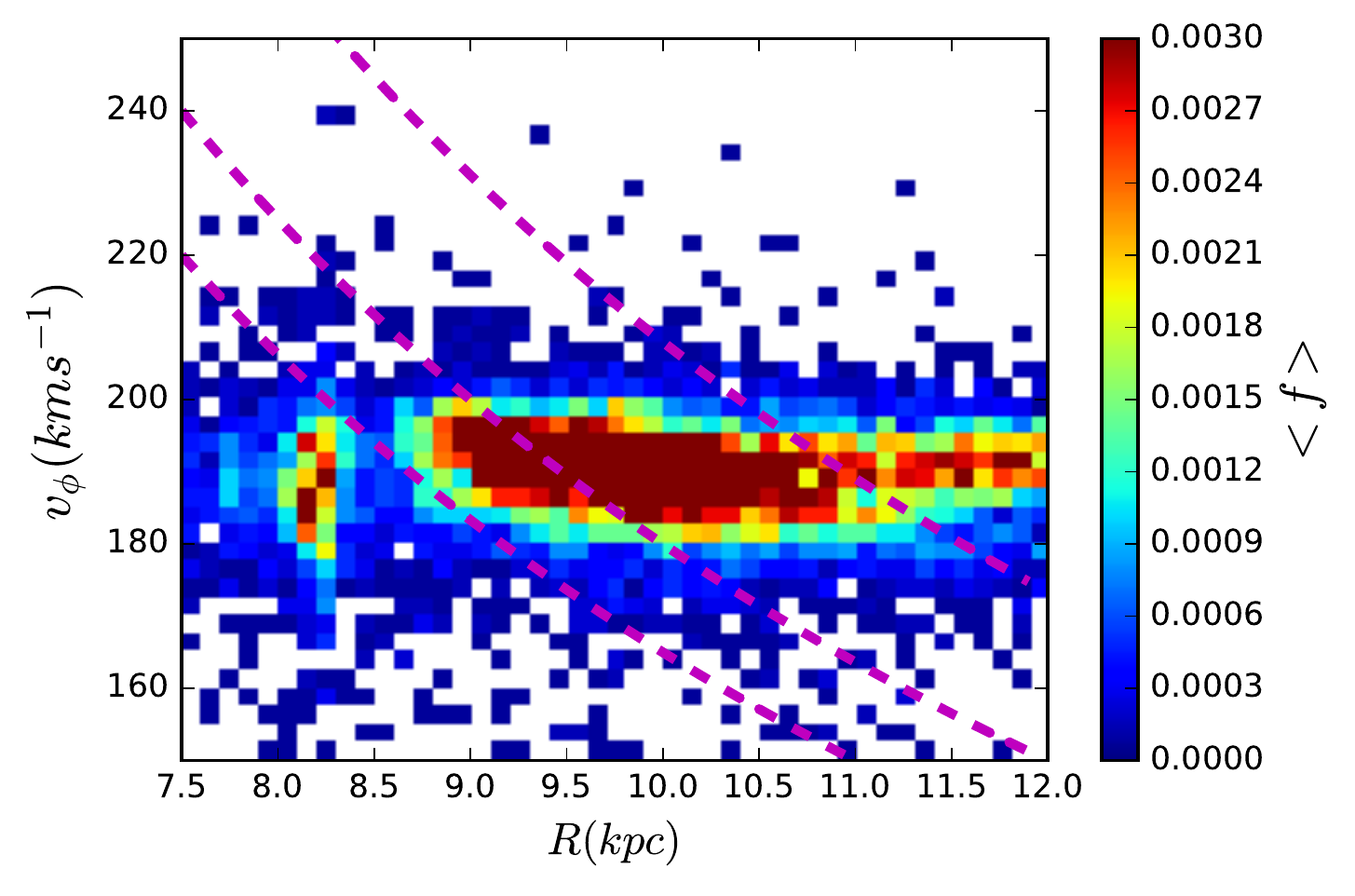}
\includegraphics[width=0.45\textwidth, trim=0.0cm 0.0cm 0.0cm 0.0cm, clip]{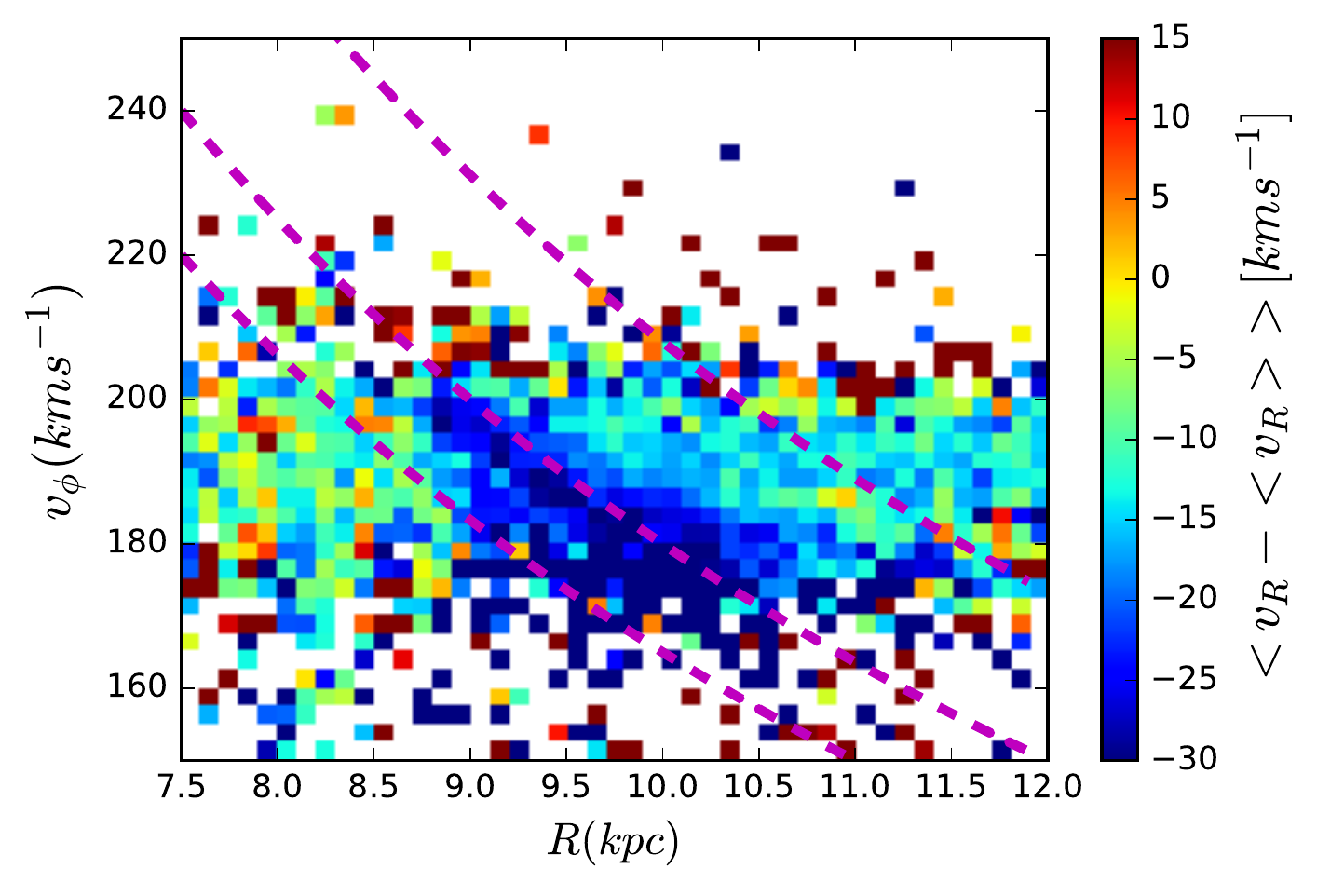}
\caption{Stars distribution in the ($R, v_{\phi}$) plane with LAMOST OB stars. Left panel is the density distribution, right panel is the mean radial velocity. The ridge pattern, especially colored by blue strips respectively in the right, is clear. The magenta dotted curves represent constant angular momentum of $L_{Z}$, similar to Fig. \ref{msden_vr_feh}.} 
\label{obden_vr}
\end{figure*}

As mentioned in the last section, the ridge pattern has been sensitive to perturbation for 0$-$14 \,Gyr. In order to know more about its population features, we make full use of LAMOST different samples. We use OB stars \citep{liu2019} to chart the distributions of density and radial velocity in the ($R, v_{\phi}$) plane, which is displayed in Fig. \ref{obden_vr}. It clearly denotes that there is an obvious ridge strip colored with blue in the right, especially for the radial velocity in the range of R from 9$-$11 \,kpc and $v_{\phi}$ from 170 to 200 km s$^{-1}$, the ridge is similar with one of the ridges in MSTO stars.  As a mater of fact, the OB stars ridges don't need to correspond to the MSTO ridges exactly due to the different population effects.

\subsection{Discussions}

As manifested and implied in \citet{wang2020b} and references therein, we suggest many mechanisms might be coupled together to cause the complexed and abundant vertical asymmetries with bending and breathing modes accompanied with mean non zero radial motions and asymmetrical rotations for the disk regions. All of these might be under a same comprehensive dynamical distribution function. In-plane asymmetries and vertical motions are coupled together as shown clearly in \citet{antoja2018, Khanna2019}, but whether other different locations and populations are still coupling together is not clear. \citet{antoja2018} used a relatively narrow range in the solar neighborhood to discover the snails in $z, v_{z}$ plane and arches, shells, box in $v_{R}, v_{\phi}, v_{z} $ plane, thus then draw the coupling conclusion, but it is not clear for ridges coupling phenomenon in $R, v_{\phi}$ plane corresponding to the larger distance range during that work. What's more, we might examine the details of the chemistry for ridges alone.

\begin{figure*}[!t]
\centering
\includegraphics[width=0.3\textwidth, trim=0.0cm 0.0cm 0.0cm 0.0cm, clip]{ridge_NS_map1.pdf}
\includegraphics[width=0.3\textwidth, trim=0.0cm 0.0cm 0.0cm 0.0cm, clip]{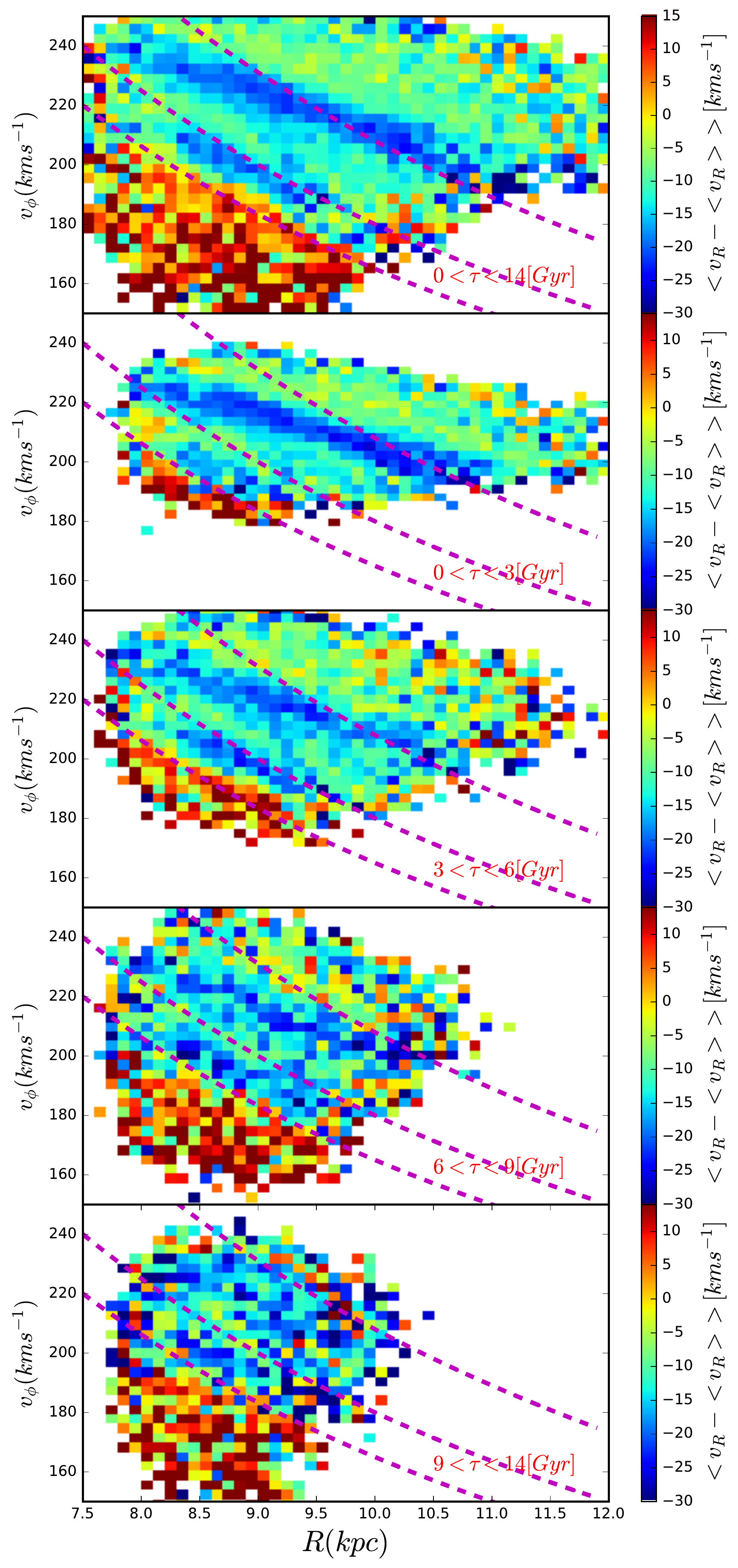}
\includegraphics[width=0.3\textwidth, trim=0.0cm 0.0cm 0.0cm 0.0cm, clip]{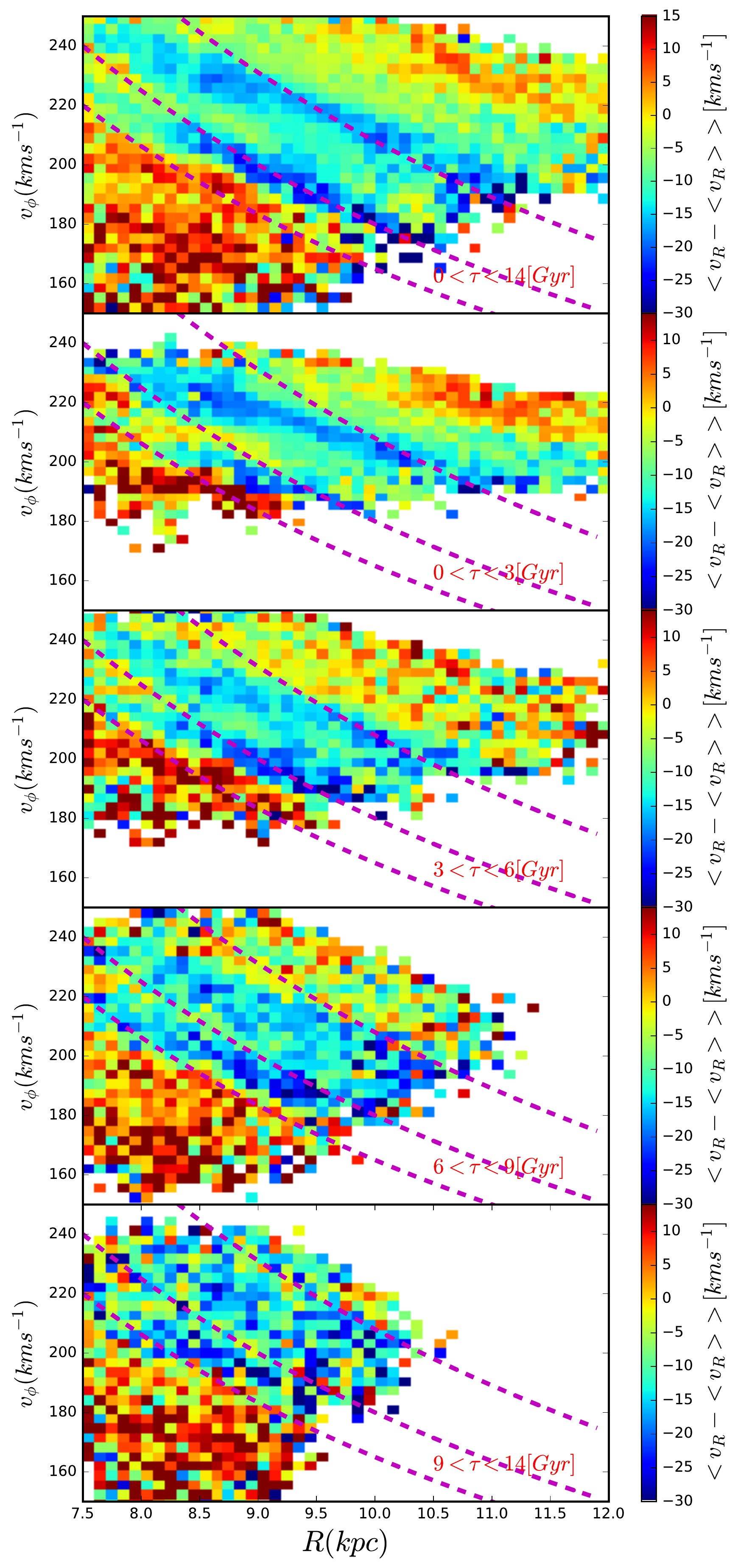}
\caption{Stars distribution in the ($R, v_{\phi}$) plane colored by radial velocity. Heat maps of various quantities are shown; Left panel is radial velocity distribution of all sample (again shown here for comparison), middle one is the southern sample, the right panel is the northern sample. The magenta dotted curves represent constant angular momentum similar to the  Fig. \ref{msden_vr_feh}.} 
\label{ridgens}
\end{figure*}

A relatively clear picture was proposed by \citet{Khanna2019}, they made use all stars of GALAH southern sky survey, test particle simulation and N-body simulation to explore the relations of ridges, arches and vertical waves, which have differences for the sky coverage and tracers with us. Here we provide the ridge pattern sensitive time to the perturbations in our sample, and suggest the angular momentum variation of ridge in different age populations and ridge distributions in the north and south side in Fig. \ref{ridgens}, etc., by using LAMOST sky survey and only MSTO and OB stars. 

\citet{Khanna2019} suggested the ridges, arches and vertical waves are coupled together, and they implied the $v_{R}$ are strongly correlated with each other and some signals are also detected in [Fe/H], [$\alpha/Fe$] and $v_{z}$, which are consistent with our main results. Meanwhile, they also pointed out clearly that phase mixing of disrupting spiral arms can generate both the ridges and arches accompanied with the points that different ridges could be originated from different scenarios in theoretical view, but if they want to unify the coupling planar and vertical motions, an intermediate satellites like Sagittarius perturbation is favored. In this work, we detect the angular momentum of one ridge pattern is relatively variable with age but other two are relatively stable, displaying the two kinds of ridges might have different origins and accompanying by the vertical signals.

We also have finished a test in Fig. \ref{ridgens}. The heat maps of various quantities show the stars radial velocity distribution in the ($R, v_{\phi}$) plane in different age populations for all sample (left), southern stars of the ridge (middle), northern stars of the ridge (right). It appears there are no clear north-south asymmetries shown here.

By investigating the origin of moving groups and diagonal ridges with the help of simulations of stellar orbits and birthplaces, \citet{Barros2020} pointed out that the diagonal ridges could be originated from the spiral resonances. There is no evidence of incomplete phase mixing in the vertical direction of the disk found in \citet{Michtchenko2019} and their results could be explained by internal mechanisms without external perturbations.  Recently, \citet{Kushniruk2020} investigated the HR 1614 moving groups and proposed that several different mechanisms such as resonances of the bar, spiral structure, phase-mixing of dissolving spiral structure, phase-mixing due to an external perturbation should be combined to explain this feature.  \citet{Laporte2020} have investigated the ridge and moving groups and found that the long bar could produce the ridge features qualitatively, meanwhile with the point that the internal and external mechanisms are shaping the Galactic disk. All these works need to explain the dependence of the ridge features on the stellar ages found in this paper. Our results provide additional constraints on the theoretical models, and encourage further theoretical studies to distinguish these scenarios based on the new observational constraints.
  
\section{Conclusion} 

In this work, using LAMOST$-$Gaia combined stars, we clearly corroborate the existences of the ridge structure in the radial velocity distribution in the $v_{\phi}, R$ plane. More importantly, with the help of three ridges detailed analysis, the evidence of the two kinds of ridge patterns with possibly different dynamical origins are firstly revealed, shown as the ridge angular momentum is relatively variable or not variable in different age populations, the two kinds of ridges are relatively clearer in the $v_{R}, L_{Z}$ plane implying again different ridges might have different physical scenarios. Moreover, the ridge patterns are also shown some features in [Fe/H], [$\alpha/Fe$], $v_{z}$ distributions.

We further investigate the kinematic analysis of the ridge pattern with different stellar ages, and find that the asymmetry is sensitive to the perturbations for 0$-$14 \,Gyr. With the help of younger populations of OB stars, we also detect ridge signals in radial velocity distribution. This is the first time stamps work on the ridge and different levels of sensitivity of different stellar populations for the response to the possible dynamical perturbation are unveiled again in this work. These features are non-trivial to be investigated in more details by us, e.g., we will go farther distance beyond 12 \,kpc to characterize it in more dimensions, which is not the target of the current work.

 \acknowledgements
We would like to thank the anonymous referee for his/her very helpful and insightful comments. This work is supported by the National Key Basic R\&D Program of China via 2019YFA0405500. H.F.W. is supported by the LAMOST Fellow project, funded by China Postdoctoral Science Foundation via grant 2019M653504 and 2020T130563, Yunnan province postdoctoral Directed culture Foundation, and the Cultivation Project for LAMOST Scientific Payoff and Research Achievement of CAMS-CAS. M.L.C. was supported by grant PGC-2018-102249-B-100 of the Spanish Ministry of Economy and Competitiveness. Y.H. acknowledges the National Natural Science Foundation of China U1531244,11833006, 11811530289, U1731108, 11803029, and 11903027 and the Yunnan University grant No.C176220100006 and C176220100007. H.W.Z. is supported by the National Natural Science Foundation of China under grant number 11973001. H.F.W. is fighting for the plan ``Mapping the Milky Way Disk Population Structures and Galactoseismology (MWDPSG) with large sky surveys" in order to establish a theoretical framework in the future to unify the global picture of the disk structures and origins with a possible comprehensive distribution function. We pay our respects to elders, colleagues and others for comments and suggestions, thanks to all of them. The Guo Shou Jing Telescope (the Large Sky Area Multi-Object Fiber Spectroscopic Telescope, LAMOST) is a National Major Scientific Project built by the Chinese Academy of Sciences. Funding for the project has been provided by the National Development and Reform Commission. LAMOST is operated and managed by National Astronomical Observatories, Chinese Academy of Sciences. This work has also made use of data from the European Space Agency (ESA) mission {\it Gaia} (\url{https://www.cosmos.esa.int/gaia}), processed by the {\it Gaia} Data Processing and Analysis Consortium (DPAC, \url{https://www.cosmos.esa.int/web/gaia/dpac/consortium}). Funding for the DPAC has been provided by national institutions, in particular the institutions participating in the {\it Gaia} Multilateral Agreement.

\end{document}